\documentclass[prc,aps,showpacs]{revtex4}
\usepackage{graphicx}
\begin{document}

\newcommand{\adag}{a^{\dag}}
\newcommand{\atil}{\tilde{a}}
\def\frp#1{${#1\over2}^+$}
\def\frm#1{${#1\over2}^-$}
\def\g{\noindent}
\def\mev{\hbox{\ MeV}}
\def\kev{\hbox{\ keV}}
\def\lambdabar{{\mathchar'26\mkern-9mu\lambda}}
\def\lambdabarrr{{^-\mkern-12mu\lambda}}

\title{Extensive R\'{e}nyi Statistics from Non-Extensive Entropy}

\author{A.S.~Parvan$^{\dag\S}$  and  T.S.~Bir\'{o}$^{\ddag}$}
\affiliation{\dag\  Bogoliubov Laboratory of Theoretical Physics,
Joint Institute for Nuclear Research, 141980 Dubna, Russia}
\affiliation{\S\ Institute of Applied Physics, Moldova Academy of
Sciences, MD-2028 Kishineu, Moldova}
\affiliation{\ddag\ MTA KFKI Research Institute for Particle and Nuclear Physics, H-1525 Budapest,
P.O.B. 49, Hungary}

\begin{abstract}
We show that starting with either the
non-extensive Tsallis entropy in Wang's formalism or the
extensive R\'{e}nyi entropy, it is possible to construct the
equilibrium statistical mechanics with non-Gibbs canonical
distribution functions. The statistical mechanics with Tsallis
entropy does not satisfy the zeroth law of thermodynamics at
dynamical and statistical independence request, whereas the extensive
R\'{e}nyi statistics fulfills all requirements of equilibrium
thermodynamics. The transformation formulas between Tsallis
statistics in Wang representation and R\'{e}nyi statistics are
presented.  The one-particle distribution
function in R\'{e}nyi statistics for classical ideal gas and finite
particle number has a power-law tail for large momenta.
\end{abstract}


\maketitle

\section{Introduction}
The Gibbs distribution function has several successful applications
in various domains of physics. It is a consequence of statistical
mechanics which satisfies all postulates of equilibrium
thermodynamics~\cite{Gibbs,Balescu}. Recently, however,  a
number of experimental data have appeared, where the asymptotic
distribution differs from the Gibbs distribution~\cite{Tsallis1}. These
deviations point out that either the equilibrium assumption
fails for high energies or there is a need to construct the
equilibrium statistical mechanics with non-Gibbs distribution
function. The Tsallis statistics explores this second alternative
on the base of a non-extensive entropy~\cite{Tsal98}. There
has been an increasing interest in studying the non-extensive
Tsallis statistics. Unfortunately, the generalized Tsallis
statistical mechanics have problems with the zeroth law of
thermodynamics. Several authors~\cite{Abe0,Wang1} have attempted to get
around this difficulty, but remain unpersuasive. Further attempts to solve
this problem were undertaken by
introducing an extensive representation for the non-extensive Tsallis
entropy~\cite{Abe1,Wang2}. As it will be shown in this work,
this idea leads to a transformation of Tsallis statistics
into the R\'{e}nyi statistics, but does not solve the problem
of second law of thermodynamics for Tsallis statistics: the
conventional correspondence between temperature and heat is less obvious. 
In this paper we point out that the statistical mechanics 
with extensive R\'{e}nyi entropy fulfill all
requirements of an equilibrium thermodynamics and still have its
canonical distribution function in a form with power-law asymptotics.
It might be, on the other hand, less preferable regarding the stability
property (H theorem based on a generalized entropy) \cite{Abe2004}.

This paper is organized as follows. In sections II and III we
discuss the microcanonical and canonical ensembles in the incomplete
non-extensive statistics and in the R\'{e}nyi statistics. The
transformation rules from Wang formalism to R\'{e}nyi statistics are
derived in section IV. In section V the properties of
thermodynamical averages and the zeroth law of thermodynamics are
discussed. Finally, in section VI
we apply these results for the classical ideal gas of massive particles.


\section{Incomplete non-extensive statistics}

The Wang's formalism of generalized statistical mechanics
uses Tsallis' alternative definition for the equilibrium
entropy~\cite{Tsal98}
\begin{equation}\label{1}
S = -k \sum\limits_i \frac{p_{i}-p_i^{q}}{1-q},
\end{equation}
and utilizes a new norm equation~\cite{Wang00,Wang1}
\begin{equation}\label{2}
    \sum\limits_{i} p_{i}^{q} = 1.
\end{equation}
Here $p_{i}^{q}$ is a probability of $i$th microscopic state of a
system and $q \in {\cal R}$ (entropic index) defines a particular
statistics. In the limit $q\to 1$ eq.~(\ref{1}) approaches the Gibbs
entropy, $S_{1}=-k\sum\limits_{i} p_{i} \ln p_{i}$. The
$q$ - expectation value of an observable $O$ in this framework
is defined as follows
\begin{equation}\label{3}
 {\cal O} = \sum\limits_{i} p_{i}^{q} \ O_{i}.
\end{equation}
From the beginning the useful function
\begin{equation}\label{4}
    \eta \equiv \sum\limits_{i} p_{i} = 1-(1-q)\frac{S}{k}
\end{equation}
should be introduced so that
\begin{equation}\label{4aa}
    S=k \ \frac{1-\eta}{1-q}.
\end{equation}
For the derivation of the distribution functions in microcanonical and canonical
ensembles we will use Jaynes' principle~\cite{J63}.


\subsection{Microcanonical ensemble $(E,V,N)$}

In order to find the distribution function $f=p_{i}^{q}$, one maximizes
the Lagrange function
\begin{equation}\label{5}
    \Phi = \frac{S}{k} - \alpha
\left(\sum\limits_{i} p_{i}^{q} - 1\right).
\end{equation}
For the probability $p_{i}^{q}$ we obtain the following
expression
\begin{equation}\label{6}
    p_{i}^{q}=\left[1-(1-q)\frac{S}{k}\right]^{\frac{q}{1-q}}.
\end{equation}
The parameter $\alpha$ has been eliminated using equation (\ref{1})
and (\ref{2}). In the limit $q\to 1$ the distribution function
(\ref{6}) has the well known form $ p_{i} = e^{-S_{1}/k}$.
Substituting eq.~(\ref{6}) into (\ref{2}) and taking into account the
conservation rules for microcanonical ensemble  results in
the equality
\begin{equation}\label{8}
    \left[1-(1-q)\frac{S}{k}\right]^{-\frac{q}{1-q}}=\sum\limits_{i} \delta_{V_{i},V}
\delta_{N_{i},N} \delta_{E_{i},E} \equiv W.
\end{equation}
Based on this we get the equipartition probability  from eq.~(\ref{6}) as a
function of the thermodynamical ensemble variables, the energy $E$,
the volume $V$ and the particle number $N$:
\begin{equation}\label{9}
    p_{i}^{q}=\frac{1}{W}
\end{equation}
The entropy is given from (\ref{6}) and (\ref{9}) by
\begin{equation}\label{10}
    S=k \frac{(W^{1/q})^{q-1}-1}{q-1}.
\end{equation}
Note that this entropy (\ref{10}) does not satisfy the famous Boltzmann
principle~\cite{Gross1}.

Differentiating  eq.~(\ref{2}) and using eqs.~(\ref{9}) and
(\ref{10}) we obtain an expression for the heat,
comprising the microcanonical limit of the second law of thermodynamics
\begin{equation}\label{11}
    \delta Q = T dS=0,
\end{equation}
where $d$ is the differential operator of three independent
variables $(E,V,N)$ and $T$ is the temperature of the system.
The first law of thermodynamics is satisfied since the heat transfer
in microcanonical ensemble is equal to zero
\begin{equation}\label{12}
  \delta Q = dE + p dV - \mu dN =0,
\end{equation}
where $p$ and $\mu$  are the pressure and chemical potential of
the system consequently. Comparing (\ref{11}) and (\ref{12}) we
obtain the fundamental equation (first law) of thermodynamics
\begin{equation}\label{13}
  T dS = dE + p dV - \mu dN.
\end{equation}
This equation is especially useful since the independent variables
are also the environmental variables.



\subsection{Canonical ensemble $(T,V,N)$}

The Lagrange function for the canonical ensemble is given by
\begin{equation}\label{1a}
\Phi = \frac{S}{k} - \alpha \left(\sum\limits_{i} p_{i}^{q} -
1\right) -\beta \left(\sum\limits_{i} p_{i}^{q} E_{i}-E\right).
\end{equation}
After maximizing the function (\ref{1a}) and using eq.~(\ref{1}) to
eliminate the parameter $\alpha$ we arrive at the following expression
for the distribution function
\begin{equation}\label{2a}
p_{i}^{q}=[1+(1-q)q\beta(\Lambda-E_{i})]^{\frac{q}{1-q}},
\end{equation}
with  $\Lambda\equiv E-S/qk\beta$. Let us now fix the
parameter $\beta$. Differentiating the function $\Lambda$ and
eq.~(\ref{1}) with respect to $\beta$, and using the distribution
function (\ref{2a}) we obtain
\begin{equation}\label{3a}
    \frac{\partial S}{\partial \beta} =k\beta \frac{\partial
    E}{\partial\beta}.
\end{equation}
So the parameter $\beta$ can be related to the temperature
\begin{equation}\label{4a}
    \frac{1}{T}\equiv \frac{\partial S}{\partial E}=
    \frac{\partial S/\partial\beta}{\partial E/\partial\beta}
    =k\beta, \;\;\;\;\;\;\;\;\;\;  \beta=\frac{1}{kT}.
\end{equation}
Thus expressing $\beta$ with the physical temperature $T$ the
distribution function (\ref{2a}) takes the form
\begin{equation}\label{5a}
p_{i}^{q}=\left[1+(1-q)\frac{\Lambda-E_{i}}{kTq^{-1}}\right]^{\frac{q}{1-q}},
\end{equation}
where  $\Lambda$ is determined from the normalization
condition eq.~(\ref{2})
\begin{equation}\label{6a}
    \sum\limits_{i} \delta_{V_{i},V} \delta_{N_{i},N} \
    \left[1+(1-q)\frac{\Lambda-E_{i}}{kTq^{-1}}\right]^{\frac{q}{1-q}}
    =1.
\end{equation}
Thus it is a function of the canonical thermodynamical variables,
$\Lambda=\Lambda(T,V,N)$. Then canonical averages are calculated
according to eq.~(\ref{3}) in the following manner
\begin{equation}\label{7a}
{\cal O} = \sum\limits_{i} \delta_{V_{i},V} \delta_{N_{i},N} \ O_{i}
    \left[1+(1-q)\frac{\Lambda-E_{i}}{kTq^{-1}}\right]^{\frac{q}{1-q}}.
\end{equation}
(For another derivation of the distribution function (\ref{5a}) see the
Appendix A.)

We would like to verify now the distribution function (\ref{5a}) from the point of
view of equilibrium thermodynamics. Applying the differential
operator of ensemble variables $(T,V,N)$ on the eqs.~(\ref{1}),
(\ref{2}) and on the energy $E$ from eq.~(\ref{3}) with eigenvalues
$E_{i}$,  and using the distribution function (\ref{5a}) and
eqs.~(\ref{2}), (\ref{3}), we obtain again the fundamental equation of
thermodynamics,
\begin{equation}\label{8a}
T dS =dE + p dV - \mu dN,
\end{equation}
where $(\partial E_{i}/\partial T)_{V,N}=0$ and the pressure and
chemical potential take the form
\begin{eqnarray}\label{9a}
  -p &=& \sum\limits_{i} p_{i}^{q} \left( \frac{\partial E_{i}}{\partial V} \right)_{T,N}, \\
  \label{10a}
  \mu &=& \sum\limits_{i} p_{i}^{q} \left( \frac{\partial E_{i}}{\partial N}
  \right)_{T,V}.
\end{eqnarray}
The free energy of the system is defined as
\begin{equation}\label{11a}
    F \equiv E - T S = \Lambda +  kTq^{-1}(1-\eta),
\end{equation}
where  $\eta$ (cf. eq.\ref{4}) is a function of the canonical
variables $(T,V,N)$:
\begin{equation}\label{12a}
    \eta =\sum\limits_{i} \delta_{V_{i},V} \delta_{N_{i},N} \
    \left[1+(1-q)\frac{\Lambda-E_{i}}{kTq^{-1}}\right]^{\frac{1}{1-q}}.
\end{equation}
The entropy is calculated from eq.~(\ref{5}). Using the definition
(\ref{11a}), from eq.~(\ref{8a}) we get
\begin{equation}\label{13a}
    dF = -S dT - p dV + \mu dN,
\end{equation}
and some further useful relations for the entropy,
pressure and chemical potential if the free energy (\ref{11a}) is
known
\begin{equation}\label{14a}
    \left(\frac{\partial F}{\partial T}\right)_{V,N} = -S,
    \;\;\;\;\;
\left(\frac{\partial F}{\partial V}\right)_{T,N} = -p, \;\;\;\;\;
\left(\frac{\partial F}{\partial N}\right)_{T,V} = \mu.
\end{equation}
The energy can be calculated from the formula
$E=-T^{2}(\partial/\partial T)(F/T)_{V,N}$.

Note that by introducing the new function $\Psi=\Psi(\Lambda)$ so
that
\begin{equation}\label{13ab}
    F \equiv -kT \frac{\Psi^{q-1}-1}{q-1},
\end{equation}
the Legendre transformation rule between entropy and free energy applies:
\begin{equation}\label{14ab}
    \frac{S}{k}=\left(\frac{\Psi^{q-1}-1}{q-1}\right) +
    T \frac{\partial}{\partial T}
    \left( \frac{\Psi^{q-1}-1}{q-1}\right).
\end{equation}

Substituting the function $\Lambda=E-q^{-1}TS$ into eq.~(\ref{5a})
we obtain the canonical distribution function in a representation
similar to a recent formalism by Tsallis~\cite{Tsal98}
\begin{equation}\label{15a}
p_{i}^{q}=\eta^{\frac{q}{1-q}}
\left[1+(1-q)\frac{E-E_{i}}{kTq^{-1}\eta}\right]^{\frac{q}{1-q}}.
\end{equation}
In this case the probability (\ref{15a}) depends on two unknown
variables $E$ and $\eta$. Thus, in order to normalize the
distribution function, one has to solve two equations. For instance
the norm equation (\ref{2}), $\sum_{i} p_{i}^{q}=1$,
\begin{equation}\label{16a}
\eta^{-\frac{q}{1-q}} =\sum\limits_{i}\delta_{V_{i},V}
\delta_{N_{i},N}
\left[1+(1-q)\frac{E-E_{i}}{kTq^{-1}\eta}\right]^{\frac{q}{1-q}}
\end{equation}
and the definition of the function $\eta=\sum_{i} p_{i}$
\begin{equation}\label{17a}
\eta^{-\frac{q}{1-q}} =\sum\limits_{i}\delta_{V_{i},V}
\delta_{N_{i},N}
\left[1+(1-q)\frac{E-E_{i}}{kTq^{-1}\eta}\right]^{\frac{1}{1-q}}.
\end{equation}
Thus $E$ and $\eta$ are functions of the canonical
thermodynamical variables, $E=E(T,V,N)$ and $\eta=\eta(T,V,N)$.
Then the entropy is calculated from formula (\ref{5}), the free
energy $F$ from eq.~(\ref{11a}). General canonical averages
(\ref{3}) take the form
\begin{equation}\label{18a}
{\cal O} = \eta^{\frac{q}{1-q}}\sum\limits_{i} \delta_{V_{i},V}
\delta_{N_{i},N} \ O_{i}
    \left[1+(1-q)\frac{E-E_{i}}{kTq^{-1}\eta}\right]^{\frac{q}{1-q}}
\end{equation}
or can be calculated from eqs.~(\ref{14a}) by differentiating the
free energy $F(T,V,N)$.


\section{R\'{e}nyi statistics}

The R\'{e}nyi statistics can be derived from the R\'enyi-entropy~\cite{Renyi1}
\begin{equation}\label{1c}
    S=k \frac{\ln(\sum_{i} p_{i}^{q})}{1-q},
\end{equation}
where $p_{i}$ is the probability to find the system in $i$th
microstate. The normalization is achieved by
\begin{equation}\label{2c}
    \sum\limits_{i} p_{i}=1
 \end{equation}
and the averages are taken as follows
\begin{equation}\label{3c}
    \mathcal{O}=\sum\limits_{i} p_{i} \mathcal{O}_{i}.
\end{equation}
Here $\mathcal{O}_{i}$ is an eigenvalue of the operator belonging to
the observable $\mathcal{O}$. The function
\begin{equation}\label{4c}
    \eta\equiv \sum\limits_{i} p_{i}^{q}=e^{(1-q)S/k},
\end{equation}
and the entropy (\ref{1c}) are simply related:
\begin{equation}\label{5c}
    S=k \frac{\ln\eta}{1-q}.
\end{equation}


\subsection{Microcanonical ensemble $(E,V,N)$}

In order to find the microcanonical distribution function one
maximizes the entropy (\ref{1c}) with an additional norm condition.
The Lagrange function becomes
\begin{equation}\label{6c}
    \Phi = \frac{S}{k} -\alpha \left(\sum\limits_{i} p_{i}-1\right).
\end{equation}
As a results, after using eq.~(\ref{4c}) to eliminate the parameter
$\alpha$, the microcanonical distribution function $p_{i}$ takes the simple Gibbs
form
\begin{equation}\label{7c}
    p_{i}=e^{-\frac{S}{k}}
\end{equation}
Then from the norm equation (\ref{2c}) we obtain
\begin{equation}\label{8c}
e^{\frac{S}{k}} = \sum\limits_{i} \delta_{V_{i},V}
\delta_{N_{i},N}\delta_{E_{i},E} \equiv W.
\end{equation}
The distribution function (\ref{7c}) and the entropy are given by
the familiar expressions
\begin{equation}\label{9c}
  p_{i} = \frac{1}{W}
\end{equation}
and
\begin{equation}
  S = k\ln W. \label{10c}
\end{equation}
The R\'{e}nyi statistics in the microcanonical ensemble resembles
the Boltzmann-Gibbs statistics.


\subsection{Canonical ensemble $(T,V,N)$}

The functional to be maximized in the canonical ensemble is given by
\begin{equation}\label{11c}
\Phi = \frac{S}{k} - \alpha \left(\sum\limits_{i} p_{i} - 1\right)
-\beta \left(\sum\limits_{i} p_{i} E_{i}-E\right).
\end{equation}
After the usual procedure the canonical distribution function takes
the power-law form
\begin{equation}\label{12c}
p_{i}=\eta^{\frac{1}{q-1}}
\left[1+(q-1)\frac{E-E_{i}}{kTq}\right]^{\frac{1}{q-1}}.
\end{equation}
In this case the probability (\ref{12c}) depends on two unknown
variables $E$ and $\eta$, and, in order to normalize the
distribution function, we solve two equations. For instance the
normalization condition (\ref{2c}), $\sum_{i} p_{i}=1$,
\begin{equation}\label{13c}
\eta^{-\frac{1}{q-1}} =\sum\limits_{i}\delta_{V_{i},V}
\delta_{N_{i},N}
\left[1+(q-1)\frac{E-E_{i}}{kTq}\right]^{\frac{1}{q-1}}
\end{equation}
and the definition of the function $\eta=\sum_{i} p_{i}^{q}$,
\begin{equation}\label{14c}
\eta^{-\frac{1}{q-1}} =\sum\limits_{i}\delta_{V_{i},V}
\delta_{N_{i},N}
\left[1+(q-1)\frac{E-E_{i}}{kTq}\right]^{\frac{q}{q-1}}.
\end{equation}
Thus $E$ and $\eta$ are functions of the canonical
thermodynamical variables, $E=E(T,V,N)$ and $\eta=\eta(T,V,N)$.
Canonical averages (\ref{3c}) take the form
\begin{equation}\label{15c}
{\cal O} = \eta^{\frac{1}{q-1}}\sum\limits_{i} \delta_{V_{i},V}
\delta_{N_{i},N} \ O_{i}
    \left[1+(q-1)\frac{E-E_{i}}{kTq}\right]^{\frac{1}{q-1}}.
\end{equation}
Eventually the entropy is calculated from the formula (\ref{5c}).

Applying the differential operator with respect to the
ensemble variables $(T,V,N)$
on eq.~(\ref{1c}) and using the distribution function (\ref{12c}),
and eqs.~(\ref{2c})--(\ref{4c}), we get the fundamental equation of
thermodynamics (\ref{8a}), $T dS =dE + p dV - \mu dN$,  and the condition
that  $(\partial E_{i}/\partial T)_{V,N}=0$. The pressure and
chemical potential are
\begin{eqnarray}\label{16c}
  -p &=& \sum\limits_{i} p_{i} \left( \frac{\partial E_{i}}{\partial V} \right)_{T,N}, \\
  \label{17cc}
  \mu &=& \sum\limits_{i} p_{i} \left( \frac{\partial E_{i}}{\partial N}
  \right)_{T,V}.
\end{eqnarray}

The free energy is given by
\begin{equation}\label{17c}
    F\equiv E-TS = E-kT\frac{\ln\eta}{1-q}
\end{equation}
and $dF = -S dT - p dV + \mu dN$. Averages can be calculated
from eqs.~(\ref{14a}) and the energy from the relation
$E=-T^{2}(\partial/\partial T)(F/T)_{V,N}$.
Note that the Legendre transformations (\ref{14ab}) are valid for
R\'{e}nyi statistics with any function $\Psi=\Psi(E,\eta)$ defined
by formula (\ref{13ab}).


\section{Transformation rules}

It is useful to express the
demands of equivalence in the microcanonical and canonical ensembles
for the distribution functions of Wang eqs.~(\ref{9}), (\ref{15a}) and
R\'{e}nyi eqs.~(\ref{9c}), (\ref{12c}) and for the averages
eq.~(\ref{3}), (\ref{3c})
\begin{eqnarray}\label{1h}
(p_{i}^{(W)})^{q^{(W)}} &=& p_{i}^{(R)}, \\
\label{2h}  \mathcal{O}^{(W)} &=& \mathcal{O}^{(R)},
\end{eqnarray}
by using the equations for the Tsallis index $q$ and temperature $T$
\begin{eqnarray}\label{24c}
  q^{(R)} &=& \frac{1}{q^{(W)}}, \\
  \label{25c}
  T^{(R)} &=& (q^{(W)})^{-1} T^{(W)} \eta^{(W)}.
\end{eqnarray}
Here the indices $W$ and $R$ refer consequently to Wang and R\'{e}nyi
statistics. Note that in the microcanonical ensemble $(E,V,N)$ we obtain
the following relations
\begin{eqnarray}\label{3h}
T^{(R)} &=& (q^{(W)})^{-1} T^{(W)} W^{\frac{q^{(W)}-1}{q^{(W)}}},\\
 \label{4h} p^{(R)} &=& p^{(W)}, \\
\label{5h}  \mu^{(R)} &=& \mu^{(W)}
\end{eqnarray}
On account of eqs.~(\ref{1h}), (\ref{24c}) the relation between
Tsallis (eq.~\ref{1}) and R\'{e}nyi (eq.~\ref{1c}) entropies
becomes
\begin{equation}\label{26c}
    S^{(R)}=\frac{k q^{(W)}}{q^{(W)}-1}\ \ln\left[1+(q^{(W)}-1)
    \frac{S^{(W)}}{k} \right].
\end{equation}
Some authors~\cite{Abe1,Wang2,Vives1} interpret this equation as
the extensive representation of Tsallis entropy, but
eqs.~(\ref{25c}) and (\ref{26c}) are transformation formulas from the Wang
formalism of Tsallis statistics to R\'{e}nyi statistics.
Substituting eqs.~(\ref{25c}) and (\ref{26c}) into (\ref{11})
results in an invariance of the second law of thermodynamics during
this transformation
\begin{equation}\label{21}
    T^{(W)} dS^{(W)} = T^{(R)} dS^{(R)}.
\end{equation}
The fundamental equation (first law) of thermodynamics eq.~(\ref{13})
is invariant, too.


\section{Properties of thermodynamical averages}

In this section we investigate
the additivity of thermodynamical averages in microcanonical and
canonical ensembles. For this reason we divide a system $C=A+B$ to
two subsystems $A$ and $B$ and make demands of dynamical and
statistical independence, i.e. we require that the energy of
microstates are additive,
\begin{equation}\label{1b}
  E_{ij}(C) = E_{i}(A) + E_{j}(B)
\end{equation}
and the probability factories
\begin{equation}\label{2b}
   p_{ij}(C) = p_{i}(A) p_{j}(B).
\end{equation}
Then the Tsallis entropy (\ref{1}) is a non-extensive variable,
following rules of pseudo-additivity~\cite{Abe0,Wang00}
\begin{equation}\label{3b}
   S(C)= S(A)+S(B)-(1-q) \frac{S(A)S(B)}{k},
\end{equation}
but the R\'{e}nyi entropy (\ref{1c}) is an extensive variable
\begin{equation}\label{1d}
   S(C)= S(A)+S(B).
\end{equation}
The average values of energy, particle number
and volume (cf. ~\ref{3}, ~\ref{3c}) are extensive variables
\begin{eqnarray}\label{4b}
  E(C) &=& E(A)+E(B), \\
  \label{5b}
  V(C) &=& V(A)+V(B), \\
  \label{6b}
  N(C) &=& N(A)+N(B),
\end{eqnarray}
under the conditions (\ref{1b})--(\ref{2b}), if they follow
restrictions on the number of particles and on the volume in the
$ij$th microstate of the system $C$:  $N_{ij}(C)=N_{i}(A)+N_{j}(C)$
and $V_{ij}(C)=V_{i}(A)+V_{j}(C)$, respectively. Substituting now
eqs.~(\ref{3b}), (\ref{4b})--(\ref{6b}) into the thermodynamical
relations following from eq.~(\ref{13}) in case of the
microcanonical ensemble $(E,V,N)$ and into the thermodynamical
relations following from eq.~(\ref{8a}) for the canonical ensemble
$(T,V,N)$, as well as using the function $\eta$ defined in
(\ref{4}), we arrive at the following relations between temperatures
in the Wang statistics:
\begin{equation}\label{7b}
    T_{C} \eta_{C}= T_{A} \eta_{A}= T_{B} \eta_{B}
\end{equation}
while the pressure and the chemical potential are equal in equilibrium:
\begin{eqnarray}
  p_{C} &=& p_{A} = p_{B},\label{8b} \\
  \mu_{C} &=& \mu_{A} =\mu_{B}. \label{9b}
\end{eqnarray}
Here $\eta_{C}=\eta_{A}\eta_{B}$ due to statistical independence
(\ref{2b}). The relation between temperatures in the R\'{e}nyi
statistics due to the extensive property of the entropy (\ref{1d})
is the conventional one:
\begin{equation}\label{7d}
    T_{C} = T_{A} = T_{B}.
\end{equation}
The pressure and the chemical potential satisfy
eq.~(\ref{8b}) and eq.~(\ref{9b}), they are intensive variables.

Thus it is proved that the non-extensive entropy of the system destroys
the intensive property of the physical temperature $T$, conjugate
to $S$. Other variables, like pressure $p$ and chemical
potential $\mu$, conjugate to  $N$ and $V$, remain intensive. The
zeroth law of thermodynamics is violated this way (see
eq.~(\ref{7b})) therefore the statistical mechanics with
non-extensive entropy (\ref{1}) does not satisfy all principles of
equilibrium thermodynamics with the probability $p_{i}^{q}$ being
the equilibrium distribution function. However, R\'{e}nyi statistics
completely satisfies all these principles.


\section{Thermodynamics of the classical ideal gas}

Finally we would like to provide an example for the comparison of
the two approaches we have discussed so far. Let us consider a
classical non-relativistic ideal gas of $N$ identical particles in
the canonical ensemble of R\'{e}nyi's and Wang's incomplete
statistics respectively. For exact evaluation of various
$q$-expectation values the integral representation of the Euler's
Gamma function will be applied, following Ref.~\cite{Prato}.

\subsection{Wang's incomplete statistics}
The functions $E$ and $\eta$ are obtained by solving
eqs.~(\ref{16a}), (\ref{17a}) for $N$ independent particles. We get
\begin{equation}
\label{1i}  E = \frac{3}{2} N kT q^{-1} \eta = \frac{3}{2} N kT_{{\rm eff}}
\end{equation}
and
\begin{equation}
 \label{2i} \eta = A^{-1} (C Z_{1}(T))^{\frac{1}{1-\frac{A}{1-q}}},
\end{equation}
where $A=1+(1-q)\frac{3}{2}N$,  $Z_{1}$ is the partition function of
the ideal Boltzmann gas in the canonical
ensemble~\cite{Huang,Parvan}, $Z_{1}=(1/N!)
(gV/\lambda_{T}^{3})^{N}$, and
$\lambda_{T}=(2\pi\hbar^{2}/mkT)^{1/2}$ is the thermal wave length
for a particle~\cite{Huang}. For the variable $A$ we have the
condition $A>0$. The variable $C$ in the case $q>1$ becomes
\begin{equation}\label{3i}
    C= \frac{\Gamma(\frac{q}{q-1}-\frac{3}{2}N)}
    {(q(q-1))^{\frac{3}{2}N} \Gamma(\frac{q}{q-1})}, \;\;\;\;\;\;\; q>1,
\end{equation}
which is valid under the condition $-3N/2+q/(q-1)>0$. In the case
$q<1$ we have
\begin{equation}\label{4i}
    C= \frac{\Gamma(\frac{1}{1-q})}
    {(q(1-q))^{\frac{3}{2}N} \Gamma(\frac{1}{1-q}+\frac{3}{2}N)}, \;\;\;\;\;\;\; q<1,
\end{equation}
where $\Gamma(z)$ is Euler's Gamma function. Then the entropy is
calculated from eq.~(5). The free energy (\ref{11a}) takes the form
\begin{equation}\label{5i}
    F=-kT \ \frac{1- \eta}{1-q} +\frac{3}{2}NkTq^{-1}\ \eta,
\end{equation}
and the pressure is obtained from eq.~(\ref{10a}):
\begin{equation}\label{6i}
    p=\frac{N}{V}kTq^{-1}\eta = \frac{2}{3} \frac{E}{V}.
\end{equation}

The $N$-particle distribution function of the classical ideal gas
in the canonical ensemble takes the following form
\begin{equation}\label{7i}
    f(\vec{p}_{1},\ldots,\vec{p}_{N})=\frac{1}{N!} \frac{(gV)^{N}}{(2\pi\hbar)^{3N}}
    \eta^{\frac{q}{1-q}} \left[1+(1-q)\frac{1}{kTq^{-1}\eta}
    (E-\sum\limits_{i=1}^{N}\frac{\vec{p}_{i}^{2}}{2m})
    \right]^{\frac{q}{1-q}} ,
\end{equation}
which is normalized to unity
\begin{equation}\label{8i}
    \int d^{3}p_{1}\ldots d^{3}p_{N} \
    f(\vec{p}_{1},\ldots,\vec{p}_{N})=1.
\end{equation}
The one-particle distribution function is defined as an integral over
the momenta of all other particles:
\begin{equation}\label{9i}
    f(\vec{p}_{1})=\int d^{3}p_{2}\ldots d^{3}p_{N} \
    f(\vec{p}_{1},\ldots,\vec{p}_{N}).
\end{equation}
Integrating formula (\ref{7i}) in Wang's incomplete statistics we
obtain
\begin{equation}\label{10i}
    f(\vec{p})=D \ \left(
    \frac{1}{2\pi mkTq^{-1}\eta A} \right)^{3/2}
    \left[1-(1-q)\frac{\vec{p}^{2}}{2mkTq^{-1}\eta A}
    \right]^{\frac{A}{1-q}-\frac{5}{2}},
\end{equation}
where the coefficient $D$ for $q>1$ is equal to
\begin{equation}\label{11i}
    D=\frac{(q-1)^{\frac{3}{2}}\Gamma(\frac{q}{q-1}-\frac{3}{2}(N-1))}
    {\Gamma(\frac{q}{q-1}-\frac{3}{2}N)}, \;\;\;\;\;\;   q>1.
\end{equation}
We have the restriction that $-3N/2+q/(q-1)>0$. In the opposite
case, for $q<1$, we have
\begin{equation}\label{12i}
    D=\frac{(1-q)^{\frac{3}{2}}\Gamma(\frac{1}{1-q}+\frac{3}{2}N)}
    {\Gamma(\frac{1}{1-q}+\frac{3}{2}(N-1))}, \;\;\;\;\;\;   q<1.
\end{equation}
The distribution (\ref{10i}) in the $N \rightarrow \infty$ limit
gives back the classical formula with an effective temperature, 
$T_{{\rm eff}}=T\eta/q$ even at finite $q-1$:
\begin{equation}\label{10x}
    f(\vec{p})= \ \left(
    \frac{1}{2\pi mkT_{{\rm eff}}} \right)^{3/2}
    e^{ - \frac{\vec{p}^{2}}{2mkT_{{\rm eff}}} }.
\end{equation}

\subsection{R\'{e}nyi statistics}
The functions $E$ and $\eta$ are found solving eqs.~(\ref{13c}),
(\ref{14c}):
\begin{equation}
\label{1g}  E = \frac{3}{2} N kT
\end{equation}
and
\begin{equation}
 \label{2g} \eta = \left(A^{-(1+\frac{qA}{1-q})} C
 Z_{1}(T)\right)^{1-q},
\end{equation}
where $A=1+(q-1)q^{-1}\frac{3}{2}N$  and it should be
$A>0$. For the variable $C$ in the case $q>1$ we have
\begin{equation}\label{3g}
    C= \frac{\Gamma(\frac{q}{q-1})}{(\frac{q-1}{q})^{\frac{3}{2}N}
\Gamma(\frac{q}{q-1}+\frac{3}{2}N)} , \;\;\;\;\;\;\; q>1,
\end{equation}
and for one in the case $q<1$ we have
\begin{equation}\label{4g}
    C= \frac{\Gamma(\frac{1}{1-q}-\frac{3}{2}N)}
    {(\frac{1-q}{q})^{\frac{3}{2}N} \Gamma(\frac{1}{1-q})}, \;\;\;\;\;\;\;
    q<1.
\end{equation}
eq.~(\ref{4g}) is valid if $-3N/2+1/(1-q)>0$. Then the entropy is
calculated from eq.~(\ref{5c})
\begin{equation}\label{5g}
    S/k=\ln Z_{1} +\ln C -(1+\frac{qA}{1-q})\ln A.
\end{equation}
The free energy (\ref{17c}) takes the form
\begin{equation}\label{6g}
    F=-kT\ln Z_{1}+\frac{3}{2}NkT-kT\ln C +(1+\frac{qA}{1-q})kT\ln A.
\end{equation}
and then the pressure is calculated from the second equation of
(\ref{10a})
\begin{equation}\label{7g}
    p=\frac{N}{V}kT = \frac{2}{3} \frac{E}{V}.
\end{equation}

The $N$-particle distribution function of classical ideal gas takes
the form
\begin{equation}\label{8g}
    f(\vec{p}_{1},\ldots,\vec{p}_{N})=\frac{1}{N!} \frac{(gV)^{N}}{(2\pi\hbar)^{3N}}
    \eta^{\frac{1}{q-1}} \left[1+(q-1)\frac{1}{kTq}
    (E-\sum\limits_{i=1}^{N}\frac{\vec{p}_{i}^{2}}{2m})
    \right]^{\frac{1}{q-1}}.
\end{equation}
Integrating formula (\ref{9i}) using eq.~(\ref{8g}) for
one-particle distribution function in R\'{e}nyi statistics we
obtain
\begin{equation}\label{9g}
    f(\vec{p})=D \ \left(
    \frac{1}{2\pi mkTqA} \right)^{3/2}
    \left[1-(q-1)\frac{\vec{p}^{2}}{2mkTqA}
    \right]^{\frac{qA}{q-1}-\frac{5}{2}},
\end{equation}
where the coefficient $D$ for $q>1$ is equal
\begin{equation}\label{10g}
    D=\frac{(q-1)^{\frac{3}{2}}\Gamma(\frac{q}{q-1}+\frac{3}{2}N)}
    {\Gamma(\frac{q}{q-1}+\frac{3}{2}(N-1))}, \;\;\;\;\;\;   q>1
\end{equation}
and in opposite case for $q<1$ is
\begin{equation}\label{11g}
    D=\frac{(1-q)^{\frac{3}{2}}\Gamma(\frac{1}{1-q}-\frac{3}{2}(N-1))}
    {\Gamma(\frac{1}{1-q}-\frac{3}{2}N)}, \;\;\;\;\;\;   q<1,
\end{equation}
which is valid in the limit $-3N/2+q/(1-q)>0$.

In this respect it should be noted that in the paper of
A.~Lavango~\cite{Lavango} the one-particle distribution function
does not correspond to the $N$-particle distribution function of
non-extensive thermostatistics.

All averages of R\'enyi statistics, eqs.~(\ref{1g})--(\ref{11g}),
can be obtained from the averages
eqs.~(\ref{1i})--(\ref{12i}) by using the transformation formulas of section
IV.

The one-particle distributions for Wang and R\'enyi statistics both
approach the Gibbs distribution in the $N\rightarrow\infty$ limit
even for $q\ne 1$. This means that non-interacting or short-range 
interacting systems, like the ideal gas, can show non-Gibbs
statistics only with finite particle number. This complies with the general
physics experience.

Fig.1 shows the one-particle distribution function for different
values of the entropic index $q$ in the above described classical
ideal gas of R\'{e}nyi statistics. The changes in behavior of the
one-particle distribution function as the number of particles grow
can be seen in Fig.2.


\section{Conclusions}

In conclusion we summarize the main idea of this paper.
Two power-law tailed 
representations for non-Gibbs canonical distribution functions $p_i=p(E_i)$
were obtained, the Wang and R\'{e}nyi statistics.
The equilibrium distribution function $p_{i}$ both for Wang and
R\'{e}nyi statistics satisfies the fundamental equation of
thermodynamics. The transformation between them
preserves the first and second laws of thermodynamics.
The same is true for the Legendre transformation
structure and for the thermodynamical relations for intensive
quantities, like pressure and chemical potential in the microcanonical
and in the canonical ensemble.

As a general rule the averages associated with R\'{e}nyi statistics
coincide with those for Wang statistics, but with a renormalized
temperature and entropy. This makes the physical interpretation of
these quantities more uncertain than it was in the Gibbs case.
The physical temperature in Wang statistics
reflects the transformation formula between Wang and R\'{e}nyi
statistics. The non-extensive entropy of Wang formalism changes, however, 
an important property of temperature, the satisfaction of the zeroth law of
thermodynamics at dynamical and statistical independence request.
The extensive R\'{e}nyi statistics, on the contrary, passes even
this test fulfilling all the requirements of equilibrium
thermodynamics. The canonical distribution function also in
this case is power-law tailed.

We have studied the R\'enyi statistics for classical ideal gas of 
massive nucleons with a finite number of particles.
The equation of state (EOS) satisfies the ideal gas law ($pV=NkT$) 
familiar from Gibbs statistics in this case, but the distribution
function differs from the Maxwell-Boltzmann one: it has a power-law tail.
This difference may be seen experimentally
only at large momenta in the tails of distributions, unless
the Tsallis index is very near to the lower limit of $q_{cr}=1-2/(3N)$
for a classical ideal gas system consisting of $N$ particles.


{\bf Acknowledgments:} This work has been supported by the Hungarian
National Research Fund OTKA (T034269) and by the Hungarian-Russian
Inter-Academy Collaboration, as well as partially supported by
the Moldavian-U.S. Bilateral Grants Program, CRDF Projects MP2-3025
and MP2-3045.  We acknowledge
valuable remarks and fruitful discussions with V.D.~Toneev and
P.~L\'{e}vai. One of authors, A.S.P, is grateful to the Department
of Theoretical Physics of KFKI for the warm hospitality during his
stay.


\appendix
\section{Distribution function of canonical ensemble in incomplete statistics}
Other definition for Lagrange function of canonical ensemble is
\begin{equation}\label{1s}
\Phi = \frac{S}{k} - \beta \left(\sum\limits_{i} p_{i}^{q}
E_{i}-E\right).
\end{equation}
After maximization of (\ref{1s}) we have
\begin{equation}\label{2s}
    p_{i}^{q}=[q(1-(1-q)\beta E_{i})]^{\frac{q}{1-q}},
\end{equation}
To normalize this distribution function it is necessary to introduce
two new parameters instead of $\beta$
\begin{equation}\label{3s}
 p_{i}^{q}=\frac{1}{Z}[1-(1-q)\beta' E_{i}]^{\frac{q}{1-q}},
\end{equation}
where the parameter $\beta'\neq\beta$. Then the parameter $Z$ is
founded from the norm eq.~(\ref{2})
\begin{equation}\label{4s}
    Z=\sum\limits_{i} \ [1-(1-q)\beta' E_{i}]^{\frac{q}{1-q}}.
\end{equation}
Let us to find the parameter $\beta'$. Substituting (\ref{3s}) into
eq.~(\ref{1}) we have
\begin{equation}\label{5s}
    S=-k\ \frac{Z^{\frac{q-1}{q}}-1}{1-q} +k\beta'Z^{\frac{q-1}{q}} \ E.
\end{equation}
Differentiating eq.~(\ref{5s}) on parameter $\beta'$ and using
eqs.~(\ref{2}), (\ref{3}) we have
\begin{equation}\label{6s}
    \frac{\partial S}{\partial \beta'}=k Z^{\frac{q-1}{q}}\left[
    \frac{\partial\ln Z^{1/q}}{\partial \beta'} (1-(1-q)\beta' E)+
    \frac{\partial}{\partial \beta'} (\beta' E)
    \right].
\end{equation}
On the other hand differentiating eq.~(\ref{1}) on $\beta'$ and
using distribution function (\ref{3s}) and eqs.~(\ref{2}), (\ref{3})
we have
\begin{equation}\label{7s}
    \frac{\partial S}{\partial \beta'}=\frac{k Z^{\frac{q-1}{q}}}{1-q}\left[
    \frac{\partial\ln Z^{1/q}}{\partial \beta'} (1-(1-q)\beta' E)+E
    \right],
\end{equation}
where was used $\partial E_{i}/\partial\beta'=0$. Comparing
eqs.~(\ref{6s}) and (\ref{7s}) the following equality is valid
\begin{equation}\label{8s}
\frac{\partial S}{\partial \beta'}=k\beta'q^{-1}Z^{\frac{q-1}{q}} \
\frac{\partial E}{\partial \beta'}.
\end{equation}
Then the parameter $\beta'$ is expressed from the temperature
\begin{equation}\label{9s}
    \frac{1}{T}\equiv \frac{\partial S}{\partial E}=
    \frac{\partial S/\partial\beta'}{\partial E/\partial\beta'}
    =k\beta'q^{-1}Z^{\frac{q-1}{q}}
\end{equation}
and
\begin{equation}\label{10s}
\beta' = \frac{q Z^{\frac{1-q}{q}}}{kT}.
\end{equation}
In this respect it should be noted that in the paper of
Q.A.~Wang~\cite{Wang00} multiplier $q$ from eq.~(\ref{10s}) was lost
as the entropy (\ref{5s}) was directly differentiated from $E$
without taking into account that the energy $E$ is not an
independent variable of canonical ensemble $(T,V,N)$.

Substituting eq.~(\ref{10s}) into eq.~(\ref{5s}) and introducing the
function $\Lambda$ in the following form (see Section II)
\begin{equation}\label{11s}
     \Lambda\equiv kTq^{-1}
     \frac{Z^{\frac{q-1}{q}}-1}{1-q}=E-q^{-1}TS
\end{equation}
for distribution function (\ref{3s}) and norm equation (\ref{4s}) we
have
\begin{equation}\label{12s}
    p_{i}^{q}=\left[1+(1-q)\frac{\Lambda-E_{i}}{kTq^{-1}} \right]^{\frac{q}{1-q}}
\end{equation}
and
\begin{equation}\label{13s}
    \sum\limits_{i} \left[1+(1-q)\frac{\Lambda-E_{i}}{kTq^{-1}}
    \right]^{\frac{q}{1-q}}=1.
\end{equation}
Thus the minimization procedure of two different functionals
(\ref{1a}) and (\ref{1s}) results in equivalency of distribution
functions (\ref{5a}) and (\ref{12s}).

\section{The classical ideal gas in Wang's representation}
\subsection{Case $q>1$}
In the case $q>1$ it is convenient to use the following integral
representation of the Euler's Gamma function~\cite{Prato},
\begin{eqnarray}
    \label{2v}
    x^{-y} &=&
\frac{1}{\Gamma(y)} \int\limits_{0}^{\infty} t^{y-1} e^{-tx} dt,
    \;\;\;\;\;\;\;\;\; Re x >0, Re y > 0.
\end{eqnarray}
To solve eqs.~(\ref{16a}), (\ref{17a}) one should substitute the
values of $x=1+(1-q)q\beta\eta^{-1}(E-E_{i})$ and $y=q/(q-1)$ in the
case of eq.~(\ref{16a}) and the values of $x$ from the above and
$y=1/(q-1)$ in the case of eq.~(\ref{17a}) into formula (\ref{2v}).
We get
\begin{equation}\label{3v}
    \eta^{\frac{q}{1-q}}=\frac{1}{\Gamma(y)}
    \int\limits_{0}^{\infty} t^{y-1} \
    e^{-t[1+(1-q)q\beta\eta^{-1}E]} \ Z_{1,N}(t(q-1)q\beta\eta^{-1}) dt,
\end{equation}
where $\beta=1/kT$ and $Z_{1,N}$ is the canonical partition function
of the ideal Boltzmann gas of $N$ identical particles with the
argument $\beta \rightarrow t(q-1)q\beta\eta^{-1}$
\begin{equation}\label{4v}
Z_{1,N}(t(q-1)q\beta\eta^{-1}) =(t(q-1)q\eta^{-1})^{-\frac{3}{2}N}
Z_{1,N}(\beta).
\end{equation}
Taking into account (\ref{2v}) and (\ref{4v}), one can carry out the
integration of integrand (\ref{3v}) with two values of $y$ from the
 eqs.~(\ref{16a}), (\ref{17a}):
\begin{eqnarray}\label{5v}
  \eta^{-\frac{q}{1-q}-\frac{3}{2}N} &=& Z_{1,N}(\beta) \
 \frac{\Gamma(\frac{q}{q-1}-\frac{3}{2}N)}
    {(q(q-1))^{\frac{3}{2}N} \Gamma(\frac{q}{q-1})} \
    [1+(1-q)q\beta\eta^{-1}E]^{\frac{q}{1-q}+\frac{3}{2}N}  \\
  \label{6v}
\eta^{-\frac{q}{1-q}-\frac{3}{2}N} &=& Z_{1,N}(\beta) \
 \frac{\Gamma(\frac{1}{q-1}-\frac{3}{2}N)}
    {(q(q-1))^{\frac{3}{2}N} \Gamma(\frac{1}{q-1})} \
    [1+(1-q)q\beta\eta^{-1}E]^{\frac{1}{1-q}+\frac{3}{2}N},
\end{eqnarray}
Solving the  eqs.~(\ref{5v}), (\ref{6v}) relative to the unknown
variables $E$ and $\eta$, we have the following expressions
\begin{eqnarray}\label{7v}
E &=& \frac{3}{2}N (q\beta)^{-1}\eta, \\
  \label{8v}
 \eta &=&  \left[1+(1-q)\frac{3}{2}N\right]^{-1} \
 \left(Z_{1,N}(\beta)
 \frac{\Gamma(\frac{q}{q-1}-\frac{3}{2}N)}
    {(q(q-1))^{\frac{3}{2}N} \Gamma(\frac{q}{q-1})}
    \right)^{-\frac{1}{\frac{q}{1-q}+\frac{3}{2}N}},
\end{eqnarray}
which are valid under the conditions $-3N/2+q/(q-1)>0$ and
$-3N/2+1/(q-1)>0$.

The one-particle distribution function are calculated from the
eqs.~(\ref{7i}), (\ref{9i}) by substituting the values of
$x=1+(1-q)q\beta\eta^{-1}(E-\sum_{i=1}^{N} \vec{p}_{i}^{2}/2m)$ and
$y=q/(q-1)$ into eq.~(\ref{2v})

\begin{equation}\label{9v}
    f(\vec{p})=\frac{gV}{N(2\pi\hbar)^{3}}
     \frac{\eta^{\frac{q}{1-q}}}{\Gamma(\frac{q}{q-1})}
    \int\limits_{0}^{\infty} t^{\frac{q}{q-1}-1} \
    e^{-t[1+(1-q)q\beta\eta^{-1}(E-\frac{\vec{p}^{2}}{2m})]} \
    Z_{1,N-1}(t(q-1)q\beta\eta^{-1}) dt.
\end{equation}
Taking into account eqs.~(\ref{2v}) and (\ref{4v}), we get for the
integrant (\ref{9v})
\begin{eqnarray}\nonumber
 f(\vec{p}) &=& \left(\frac{\beta}{2\pi m}\right)^{3/2}
 \eta^{\frac{q}{1-q}+\frac{3}{2}(N-1)} Z_{1,N}(\beta) \
 \frac{\Gamma(\frac{q}{q-1}-\frac{3}{2}(N-1))}
    {(q(q-1))^{\frac{3}{2}(N-1)} \Gamma(\frac{q}{q-1})} \times \\ \label{10v}
 && \times \left[1+(1-q)q\beta\eta^{-1}(E-\frac{\vec{p}^{2}}{2m})
    \right]^{\frac{q}{1-q}+\frac{3}{2}(N-1)}.
\end{eqnarray}
Substituting the eqs.~(\ref{7v}) and (\ref{8v}) into expression
(\ref{10v}) we obtain the one-particle distribution function
(\ref{10i}) with the coefficient (\ref{11i}) under restrictions
described in the main text.

\subsection{Case $q<1$}
In the case $q<1$ should be used the Hankel's contour integral in
the complex plane~\cite{Abram} making the transformation
$t\rightarrow tx$ with $x$ real and positive~\cite{Prato},
\begin{eqnarray}
    \label{12v}
    x^{y-1} &=&
\Gamma(y) \frac{\imath}{2\pi}\oint\limits_{C}
 (-t)^{-y} e^{-tx} dt,
    \;\;\;\;\;\;\;\;\;  Re x >0,   | y |<\infty.
\end{eqnarray}
To solve eqs.~(\ref{16a}), (\ref{17a}) one should substitute the
values of $x=1+(1-q)q\beta\eta^{-1}(E-E_{i})$ and $y=1/(1-q)$ in the
case of eq.~(\ref{16a}) and the values of $x$ from the above and
$y=(2-q)/(1-q)$ in the case of eq.~(\ref{17a}) into formula
(\ref{12v}). We get
\begin{equation}\label{13v}
    \eta^{-\frac{q}{1-q}}=\Gamma(y) \frac{\imath}{2\pi}\oint\limits_{C}
 (-t)^{-y}
       e^{-t[1+(1-q)q\beta\eta^{-1}E]} \ Z_{1,N}(t(q-1)q\beta\eta^{-1}) dt,
\end{equation}
After calculations similar to eqs.~(\ref{3v})--(\ref{8v}) for the
energy $E$ and the function $\eta$, we have
\begin{eqnarray}\label{17v}
E &=& \frac{3}{2}N (q\beta)^{-1}\eta, \\
  \label{18v}
 \eta &=&  \left[1+(1-q)\frac{3}{2}N\right]^{-1} \
 \left(Z_{1,N}(\beta)
 \frac{(q(1-q))^{-\frac{3}{2}N} \Gamma(\frac{1}{1-q})}
 {\Gamma(\frac{1}{1-q}+\frac{3}{2}N)}
    \right)^{-\frac{1}{\frac{q}{1-q}+\frac{3}{2}N}}.
\end{eqnarray}

The one-particle distribution function are calculated from the
eqs.~(\ref{7i}), (\ref{9i}) by substituting the values of
$x=1+(1-q)q\beta\eta^{-1}(E-\sum_{i=1}^{N} \vec{p}_{i}^{2}/2m)$ and
$y=1/(1-q)$ into eq.~(\ref{12v})
\begin{equation}\label{19v}
    f(\vec{p})=\frac{gV}{N(2\pi\hbar)^{3}}
     \eta^{\frac{q}{1-q}}\Gamma(\frac{1}{1-q})
     \frac{\imath}{2\pi}\oint\limits_{C}
 (-t)^{-\frac{1}{1-q}} \
    e^{-t[1+(1-q)q\beta\eta^{-1}(E-\frac{\vec{p}^{2}}{2m})]} \
    Z_{1,N-1}(t(q-1)q\beta\eta^{-1}) dt.
\end{equation}
Taking into account eqs.~(\ref{12v}) and (\ref{4v}), we get for the
integrant (\ref{19v})
\begin{eqnarray}\nonumber
 f(\vec{p}) &=& \left(\frac{\beta}{2\pi m}\right)^{3/2}
 \eta^{\frac{q}{1-q}+\frac{3}{2}(N-1)} Z_{1,N}(\beta) \
 \frac{(q(1-q))^{-\frac{3}{2}(N-1)} \Gamma(\frac{1}{1-q})}
 {\Gamma(\frac{1}{1-q}+\frac{3}{2}(N-1))}
     \times \\ \label{20v}
 && \times \left[1+(1-q)q\beta\eta^{-1}(E-\frac{\vec{p}^{2}}{2m})
    \right]^{\frac{q}{1-q}+\frac{3}{2}(N-1)}.
\end{eqnarray}
Substituting the eqs.~(\ref{17v}) and (\ref{18v}) into expression
(\ref{20v}) we obtain the one-particle distribution function
(\ref{10i}) with the coefficient (\ref{12i}).

\newpage
\begin{figure}
  \includegraphics[width=100mm,angle=-90]{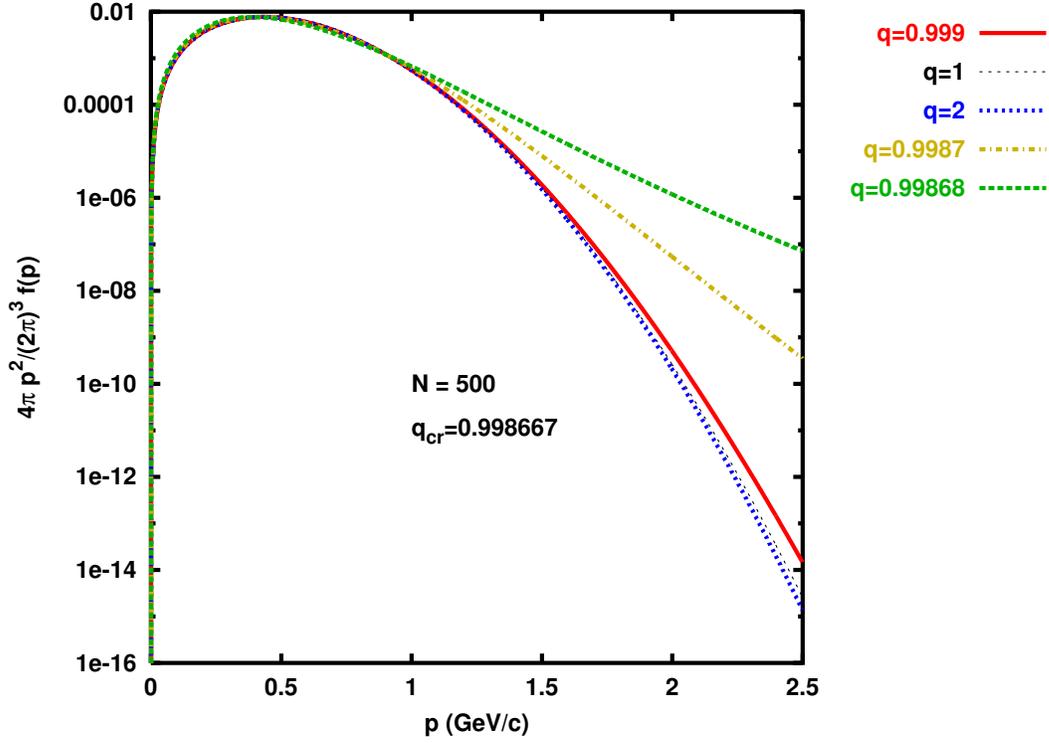}\\
  \caption{The one-particle distribution function for classical ideal
  gas of $N=500$ identical nucleons in extensive R\'{e}nyi statistics
  at the temperature $T=100$ MeV for different values of $q$: $q=1$ 
  (the second line from below) 
  correspond to the Maxwell's distribution function, other lines
  to the R\'{e}nyi distribution function.}\label{fig1}
\end{figure}

\begin{figure}
  \includegraphics[width=100mm,angle=-90]{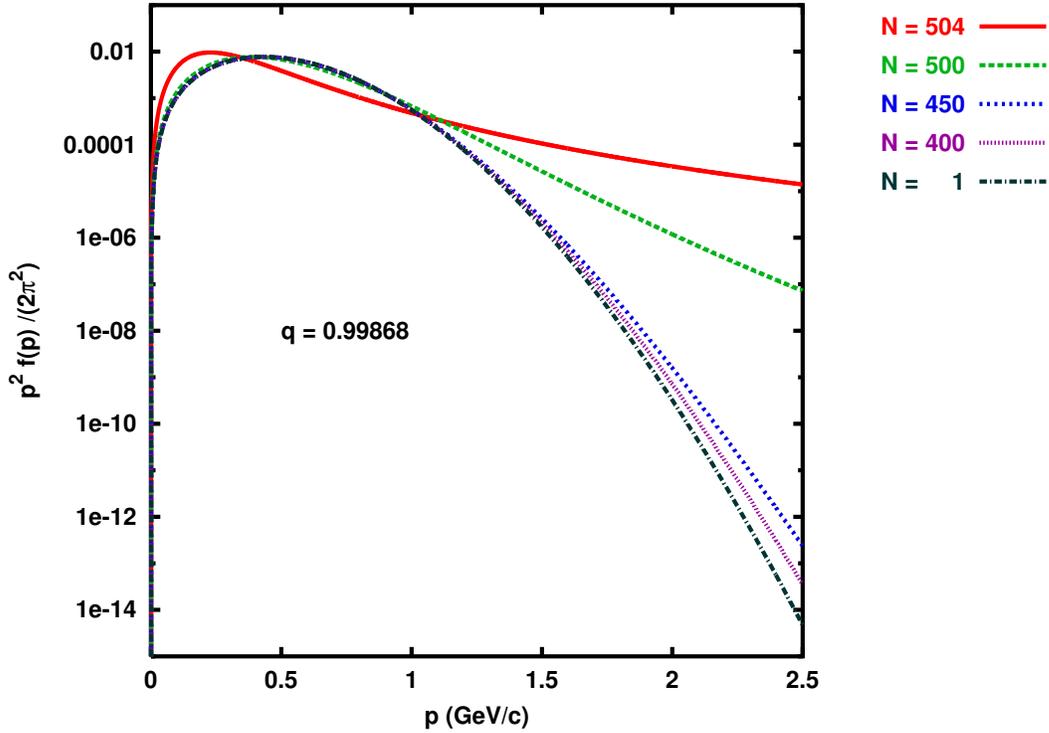}\\
  \caption{The one-particle distribution functions for classical ideal gas of
  $N=504, 500, 450, 400$ and $1$ nucleons from the top down 
  in extensive R\'{e}nyi statistics at the temperature $T=100$ MeV 
  and $q=0.99868$. The value of $N$ is restricted to 
  $N < \frac{2}{3}\frac{q}{1-q \approx 504.38}$ at this choice of the 
  Tsallis index.
  }\label{fig2}
\end{figure}


\end{document}